\documentclass[a4paper,11pt]{article}
\usepackage{pos}
\usepackage{booktabs}
\usepackage{tikz}
\usetikzlibrary{arrows.meta,positioning}

\title{Learning Low Modes of the Wilson Dirac Operator\\
with Gauge-Equivariant Networks}
\ShortTitle{Low Modes with Gauge-Equivariant Networks}

\author*[a]{Minjae Choi}
\author[b]{Hiroshi Ohno}

\affiliation[a]{Graduate School of Science and Technology,\\
  University of Tsukuba, Tsukuba, Ibaraki 305-8571, Japan}
\affiliation[b]{Center for Computational Sciences,\\
  University of Tsukuba, Tsukuba, Ibaraki 305-8577, Japan}

\emailAdd{choi@het.ph.tsukuba.ac.jp}
\emailAdd{hohno@ccs.tsukuba.ac.jp}

\abstract{
In adaptive multigrid, test vectors that better approximate the low-mode subspace of the Wilson Dirac operator $D$ can reduce the solve time, but their construction can be costly. In some cases, the setup takes longer than a single solve.  To reduce the cost of generating test vectors, we train
networks built from gauge-equivariant parallel-transport layers once on a
quenched ensemble, using a Rayleigh-Ritz loss for $D^\dagger D$. We then test whether the trained networks maintain their performance at other hopping parameters and on configurations not used in training. In these tests, the network outputs are used as test vectors, and the costly $V$-cycle refinement in the adaptive setup is omitted to reduce the setup time. This substantially reduces the setup time while leaving the solve time largely unchanged in most cases. Near the critical hopping parameter $\kappa_c$, the solve time is also reduced in some cases.
}

\FullConference{
The 43rd International Symposium on Lattice Field Theory (Lattice 2026)\\
University of Maryland, College Park, USA\\
26 July -- 1 August 2026
}

\begin{document}
\maketitle

\section{Introduction}

Lattice QCD (LQCD) requires enormous computational costs, and one of its main bottlenecks is solving the linear
equation $Du=b$, where $D$ is the Wilson Dirac operator, $b$ a given source,
and $u$ the fermion field to be computed. 
At fixed lattice spacing and quark mass, the computational cost increases
with the lattice volume because a larger volume entails both a larger
linear system and more low modes~\cite{Luscher:2007}.
Low-mode components of the error hinder the convergence of conventional
iterative solvers, particularly at small quark masses~\cite{Frommer:2013}.
Algebraic multigrid (AMG) addresses this difficulty by treating these components
on a coarse grid~\cite{Brannick:2008,Babich:2010}.
The effectiveness of adaptive multigrid depends on how well the coarse space spanned by its test vectors captures the low-mode components of the error.
Since this space depends on the gauge configuration, it generally needs to be rebuilt or updated when the configuration changes, adding to the overall computational cost.

In this work, we use a gauge-equivariant neural network to construct test
vectors, aiming to reduce the computational cost relative to the conventional
adaptive setup. Such networks have previously been explored for
preconditioning~\cite{Lehner:2023,Pfahler:2026} and multigrid
setup~\cite{Lehner:2024,Knuttel:2024}. Our aim is to learn features that
generalize across gauge configurations, allowing the network to generate
effective test vectors on new configurations.
We train the network layers by minimizing the Ritz values of $D^\dagger D$, computed using the Rayleigh-Ritz procedure in the subspace spanned by its output vectors. Through this optimization, the network learns to generate vectors spanning an approximate 
low-mode subspace of $D^\dagger D$. 

As a first application, we use the network-generated vectors as test vectors in domain
decomposition adaptive AMG (DD-$\alpha$AMG)~\cite{Frommer:2013} for the Wilson Dirac operator. In its
standard setup, random test vectors are smoothed and iteratively refined
through $V$-cycles for each configuration. To evaluate the effectiveness of the generated vectors as test vectors, we replace the random vectors with the network outputs, retain the initial SAP smoothing and aggregation steps, and skip the iterative $V$-cycle refinement of the adaptive setup. This reduces the per-configuration setup time (after a one-time training cost amortized across the fixed-volume $16^{3} \times 32$ ensemble) for parameter sets tested in this study, while the solve time stays comparable or becomes shorter near the critical hopping parameter $\kappa_c$ with a sufficient number of test vectors.

\section{Architecture and Training}

\subsection{Parallel-transport layer}
\label{sec:pt}

Our architecture adopts the parallel-transport (PT) and linear layers introduced by \cite{Pfahler:2026}. The PT layer is based on the hopping operators \(H_{\pm\mu}\), which act on a fermion field \(\phi\) as
\[
(H_{+\mu}\phi)(x)=U_\mu^\dagger(x-\hat\mu)\,\phi(x-\hat\mu),
\qquad
(H_{-\mu}\phi)(x)=U_\mu(x)\,\phi(x+\hat\mu),
\]
where \(x\) denotes a lattice site, \(\hat\mu\) the unit vector in direction \(\mu=1,\ldots,4\), and \(U_\mu(x)\in\mathrm{SU}(3)\) the gauge link from \(x\) to \(x+\hat\mu\). Let \(\mathcal{P}\) denote a fixed set of lattice paths, and let \(T_p[U]\) denote the parallel-transport operator along a path \(p=(s_1,\ldots,s_\ell)\in\mathcal{P}\), defined by
\[
T_p[U]=H_{s_\ell}\cdots H_{s_1},
\]
where \(s_i=\pm\mu\) denotes the \(i\)-th step and \(\ell\) is the number of hopping operators in \(T_p[U]\). A path with
\(s_1=\cdots=s_\ell=\pm\mu\) is a straight path in direction
\(\pm\mu\), and the empty path \(p=()\) with \(\ell=0\) gives \(T_p[U]=1\).

A PT layer maps \(n\) input fermion fields \(\phi_1,\ldots,\phi_n\), together with the gauge field \(U\), to \(m\) output fields \(\psi_1,\ldots,\psi_m\) as 
\[
\psi_a(x)
=\sum_{b=1}^{n}
W_{ab}
\bigl(T_{p^{(b)}}[U]\phi_b\bigr)(x),
\qquad a=1,\ldots,m,
\]
where $W_{ab} \in \mathbb{C}^{4\times 4}$ are trainable spin-mixing matrices and $T_{p^{(b)}}[U]$ parallel transports $\phi_b$ along the fixed path $p^{(b)}$. Here, we set $n=\vert \mathcal{P}\vert$ and assign each input field a distinct path $p^{(b)}$ in $\mathcal{P}$, so that every path is used exactly once. The PT layer is gauge equivariant, provided that both the gauge field and the input fermion fields transform under the same local gauge transformation \(G(x)\in\mathrm{SU}(3)\). In particular, if \(U_\mu(x)\to U^G_\mu(x)=G(x)\,U_\mu(x)\,G(x+\hat\mu)^\dagger\) and \(\phi_b(x)\to\phi_b^G(x)=G(x)\phi_b(x)\), then
\[
\bigl(T_{p^{(b)}}[U^G]\phi_b^G\bigr)(x)
=
G(x)\bigl(T_{p^{(b)}}[U]\phi_b\bigr)(x).
\]
Since the spin-mixing matrices act independently of the color degrees of freedom, the output fields transform as
\[
\psi_a^G(x)=G(x)\psi_a(x).
\]
We use straight paths with \(\ell=1,2,4,\) and \(5\)
in both the positive and negative directions along each of the four
space-time axes. Together with the empty path \(p=()\), which leaves the field
unchanged, these give a total of 33 paths,
\[
\mathcal{P}=\{()\}\cup\bigl\{p=(s_1,\ldots,s_\ell)\;:\;s_1=\cdots=s_\ell=\pm\mu,\ \mu=1,\ldots,4,\ \ell\in\{1,2,4,5\}\bigr\}.
\]
In general, each PT layer can use its own set of paths; for instance, the first PT layer
could use only straight paths and the second only single-hop paths (\(\ell=1\)).
In this work, however, we use the same set \(\mathcal{P}\) in every PT layer.

\subsection{Network architecture}

\begin{figure}[!hb]
\centering
\begin{tikzpicture}[
  font=\small,
  >=Latex,
  box/.style={draw=black!70, rounded corners=2pt, align=center,
              inner sep=3pt, minimum height=1.7em},
  pt/.style={box, fill=blue!12, minimum width=1.7cm},
  local/.style={box, fill=white, minimum width=1.5cm},
  io/.style={box, fill=green!12, minimum width=2.0cm},
  result/.style={box, fill=red!12, minimum width=1.9cm},
  arr/.style={->, thick, black!70},
  gauge/.style={->, thick, dashed, green!50!black},
  note/.style={font=\footnotesize, align=center},
]
  \node[local] (ff)
        {FermionField\\layer};
  \node[local, right=0.28cm of ff] (dense)
        {Linear\\layer};
  \node[pt, right=0.28cm of dense] (ptstack)
        {PT layer\\$\times 32$};
  \node[local, right=0.28cm of ptstack] (bs2)
        {BlockScalar\\layer};
  \node[local, right=0.28cm of bs2] (norm)
        {Norm\\layer};
  \node[result, right=0.28cm of norm] (eig)
        {$K$ output\\vector fields};

  \node[io, above=0.6cm of ptstack] (U)
        {input: $\{U_\mu\}$};

  \node[local, below=0.6cm of eig] (ritz)
        {RR layer\\{\footnotesize training only}};
  \node[result, right=0.28cm of ritz] (loss)
        {scalar loss\\$\mathcal{L}_{\text{RR}}$};

  \draw[arr] (ff) -- (dense);
  \draw[arr] (dense) -- (ptstack);
  \draw[arr] (ptstack) -- (bs2);
  \draw[arr] (bs2) -- (norm);
  \draw[arr] (norm) -- (eig);
  \draw[arr] (eig) -- (ritz);
  \draw[arr] (ritz) -- (loss);

  \draw[gauge] (U) -- (ptstack)
    node[midway, right, note, text=green!50!black]
    {parallel transport};

  \node[note, font=\normalsize, text=black!60, below=0.15cm of dense, text width=1.9cm]
        {expand vectors};
  \node[note, font=\normalsize, text=black!60, below=0.15cm of ptstack, text width=2.1cm]
        {GE feature extract};
  \node[note, font=\normalsize, text=black!60, below=0.15cm of bs2, text width=1.9cm]
        {block rescale};
\end{tikzpicture}%
\caption{Overall network architecture. Trainable fermion fields are processed
by Dense, PT~$\times 32$, BlockScalar, and Norm layers to produce $K$ output
vectors. The gauge field enters only through the PT stack. During training the
outputs feed a Rayleigh-Ritz loss on $D^\dagger D$; after training they are
used as multigrid test vectors.}
\label{fig:network}
\end{figure}
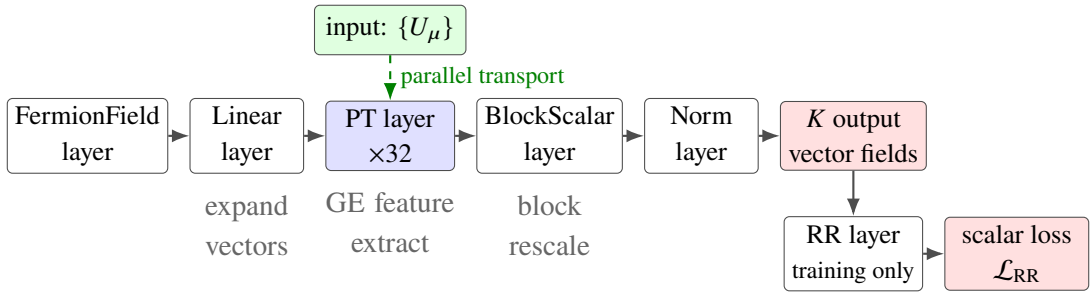

Figure~\ref{fig:network} shows the network used in this study. It maps the gauge field $U$
to $K$ fermion fields $v_1,\ldots,v_K$, which are trained to span an approximated low-mode subspace of $D^\dagger D$; after training, the same fields serve as the $K$ test
vectors of the multigrid setup. We describe the layers in the order in which
they are applied. Each layer maps its input fields $\phi$ to output fields $\psi$
and has its own trainable weights. The outputs of one layer serve as the inputs to the next.

\paragraph{FermionField layer}
The FermionField layer holds a single fermion field $\phi$ 
with $\phi(x)\in\mathbb{C}^{4\times N_c}$, corresponding to four spin and $N_c=3$ color components, at each lattice site $x$. All of these components are
trainable parameters, and $\phi$ does not depend on $U$.
After training, $\phi$ is held fixed and does not
automatically transform when $U$ is gauge transformed. Since all subsequent layers are gauge equivariant, gauge equivariance of the full network is preserved by explicitly transforming $\phi$ together with $U$. Writing the output as $v_k[U,\phi]$, one then has
$v_k[U^G,G\phi]=G\,v_k[U,\phi]$.

\paragraph{Linear layer}
A linear layer maps \(n\) input fields \(\phi_1,\ldots,\phi_n\) to \(m\)
output fields by on-site spin mixing,
\[
\psi_a(x)=\sum_{b=1}^{n}W_{ab}\,\phi_b(x),
\qquad a=1,\ldots,m,
\]
where the trainable matrices \(W_{ab}\in\mathbb{C}^{4\times4}\) act on the spin
index. It has the form of the PT-layer equation with every \(T_{p^{(b)}}[U]\)
replaced by the identity. In our network, it maps the single trainable field
\(\phi\) (\(n=1\)) to \(m=|\mathcal{P}|=33\) fields, one for each path of the
first PT layer.

\paragraph{PT layers}
Next come 32 PT layers as defined in Sec.~\ref{sec:pt}. The first 31 map
$|\mathcal{P}|$ fields to $|\mathcal{P}|$ fields, one for each path of the
next layer, and the last has $m=K$ outputs $\psi_1,\ldots,\psi_K$. There is
no nonlinear activation function between the layers, so $\psi_k$ depends
linearly on the trainable field $\phi$.

\paragraph{Block-scaling layer}
The block-scaling layer (BlockScalar) takes these $K$ fields as its input
$\phi_k$ and rescales each of them by a trainable complex number that is
constant on lattice blocks,
\[
\psi_k(x)=w_{B_r(x),k}\,\phi_k(x),
\qquad
w_{B,k}\in\mathbb{C},
\qquad
k=1,\ldots,K,
\]
where the lattice is divided into non-overlapping hypercubic blocks of side
length $r$, i.e., $r^4$ sites each, $B_r(x)$ is the block that contains $x$,
and $w_{B,k}$ is the weight of block $B$ for field $k$. We use $r=2$, whereas
DD-$\alpha$AMG~\cite{Frommer:2013} divides the lattice into aggregates of
side length $4$, restricts the test vectors to each aggregate and
orthonormalizes them there to form the prolongator (Sec.~\ref{sec:mg}). Each
aggregate thus consists of $2^4=16$ blocks. If the two block sizes were equal, the layer
would only rescale each test vector by a constant within each aggregate and
would therefore leave the local subspace used to construct the prolongator
unchanged. Choosing $r=2$ allows the layer to modify the variation of each
vector within an aggregate. Since \(w_{B,k}\) is a complex number, it commutes
with \(G(x)\), so \(\phi_k^G(x)=G(x)\phi_k(x)\) implies
\(\psi_k^G(x)=G(x)\psi_k(x)\); the layer is therefore gauge equivariant.

\paragraph{Normalization layer}
Finally, the normalization layer (Norm) normalizes each of its input fields
$\phi_k$ over the whole lattice,
\[
v_k=\frac{\phi_k}{\lVert\phi_k\rVert},
\qquad
\lVert\phi_k\rVert^2=\sum_x\operatorname{tr}\bigl[\phi_k(x)^\dagger\phi_k(x)\bigr],
\]
where the sum runs over all lattice sites and the trace over spin and color.
Since $\lVert\phi_k\rVert$ is gauge invariant, this layer is also gauge
equivariant. Unlike batch normalization, it has no regularizing effect; it only removes the
overall scale of each field, which keeps the values in a stable numerical range
through the linear layers.

\subsection{Rayleigh-Ritz loss}
\label{sec:ritz}

Adaptive multigrid benefits from test vectors $v$ that $D$ nearly annihilates, i.e., vectors
with small $\lVert Dv\rVert/\lVert v\rVert$. Since $D$ is not Hermitian, its
Rayleigh quotient $v^\dagger Dv/v^\dagger v$ is complex and cannot be
minimized. We therefore train the
network with a Rayleigh-Ritz loss on the Hermitian, positive semi-definite
operator $D^\dagger D$. Its Rayleigh quotient is
$v^\dagger D^\dagger D v/v^\dagger v=\lVert Dv\rVert^2/\lVert v\rVert^2$, and
its low modes are the right singular vectors of $D$ with the smallest
singular values. Since $D$ satisfies $\gamma_5 D\gamma_5=D^\dagger$ with the
Dirac matrix $\gamma_5$, one has $D^\dagger D=(\gamma_5 D)^2$, so the low-mode
subspace of $D^\dagger D$ is also spanned by the eigenvectors of the Hermitian
operator $\gamma_5 D$ with eigenvalues of smallest magnitude. We do not use $\gamma_5 D$ directly because it is
indefinite, and minimizing its Rayleigh quotient would target its most
negative eigenvalues instead.

Let \(V\in\mathbb{C}^{N\times K}\) collect the \(K\) network output vectors as columns, where \(N\) is the number of complex components of a fermion field on the lattice,
\[
V=\begin{bmatrix}v_1&\cdots&v_K\end{bmatrix}.
\]
Writing \(A[U]\equiv D[U]^\dagger D[U]\), the generalized Ritz values \(\lambda_k\) are the eigenvalues of the \(K\times K\) problem
\[
\hat{A}=V^\dagger A[U]\,V,
\qquad
S=V^\dagger V,
\qquad
\hat{A}c_k=\lambda_k S c_k,
\]
where \(c_k\in\mathbb{C}^K\) are the generalized eigenvectors.
The training loss is the sum of these Ritz values,
\[
\mathcal{L}_{\text{RR}}
=\sum_{k=1}^{K}\lambda_k
=\operatorname{tr}(S^{-1}\hat{A})
=\operatorname{tr}\!\left[
(V^\dagger V)^{-1}V^\dagger A[U]\,V
\right].
\]
Minimizing \(\mathcal{L}_{\text{RR}}\) pushes the columns of \(V\) toward a low-mode subspace of \(A[U]\). The loss depends on \(V\) only through this \(K\times K\) problem. A trace objective of the same form is used by spectral inference networks to learn eigenfunctions of linear operators~\cite{Pfau:2019}. The Rayleigh-Ritz problem is solved only during training. After training, the columns $v_1,\ldots,v_K$ of $V$ are used as
the initial test vectors for the multigrid setup where SAP smoothing
is applied while the adaptive $V$-cycle refinement is omitted.

\subsection{Training setup}

We train the model on quenched gauge configurations generated with the heatbath algorithm at gauge coupling \(\beta=6.0\) on a \(16^3\times32\) lattice. The Wilson Dirac operator is evaluated at \(\kappa=0.158\). This value was chosen based on preliminary tests of the dependence of the iterative-solver performance on \(\kappa\), which showed that the benefit of the learned vectors becomes more pronounced as the system approaches the critical-mass regime.
We train separate models with \(K\in\{12,24,48,64\}\) output vectors, where \(K\) determines the dimension of the learned test-vector subspace and, consequently, the size of the corresponding coarse space. The network parameters are initialized using Xavier initialization and optimized with Adam.
Our implementation is written in Julia and builds on the packages
\texttt{Gaugefields.jl} and \texttt{LatticeDiracOperators.jl}~\cite{Nagai:2024,Nagai:2026}.
Each model is trained for 50 epochs on 320 gauge configurations. An additional 20 configurations from the same ensemble, which are not used for training, are reserved as a test set, and the Rayleigh-Ritz loss on these configurations is monitored during training.

\begin{figure}[t]
\centering
\includegraphics[width=\textwidth]{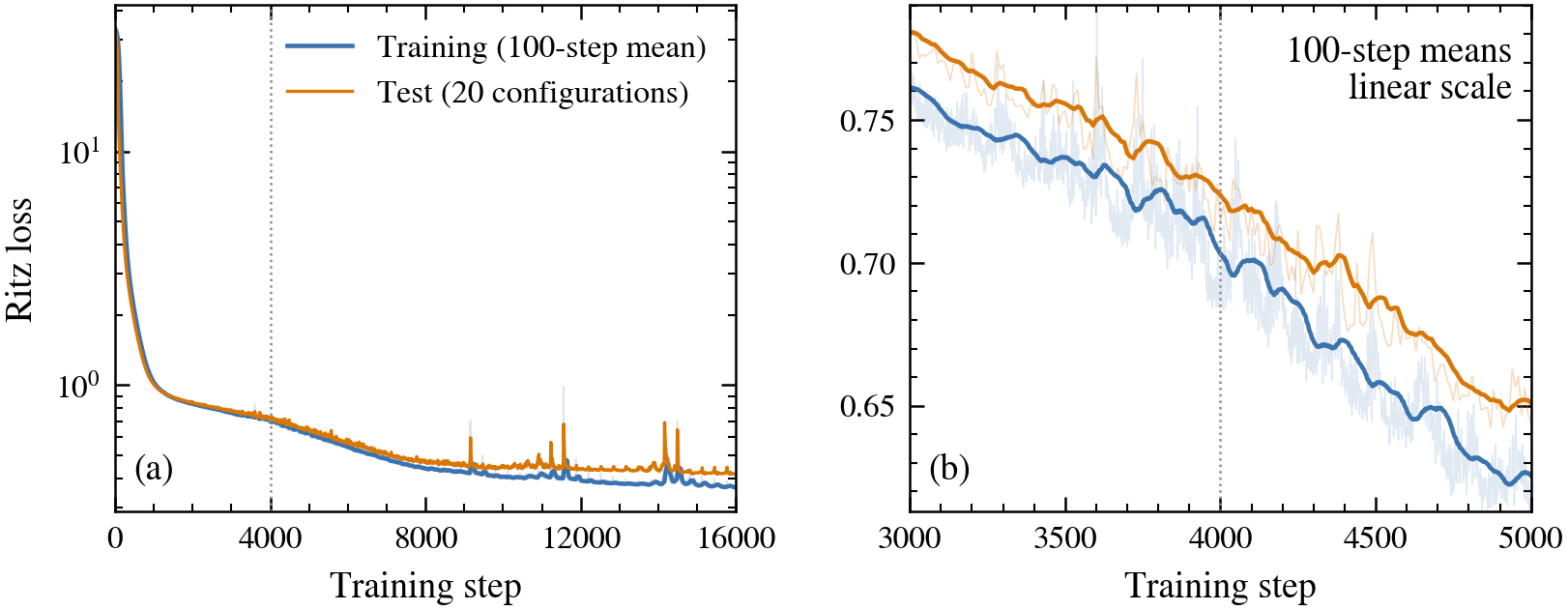}
\caption{Rayleigh-Ritz loss of the \(K{=}24\) model at \(\kappa{=}0.158\) on the 320
training configurations (blue) and on the 20 held-out test configurations
(orange). (a) Full training history on a logarithmic scale; the training
loss is shown as a trailing mean over 100 steps, and the test loss is
evaluated every 10 steps. The 16000 training steps correspond to 50 epochs.
(b) Losses from 3000 to 5000 steps on a linear scale. Thick lines are
100-step means, and thin lines show the unsmoothed values. In both panels,
the vertical dotted line marks the checkpoint at 4000 steps used in this work.}
\label{fig:training}
\end{figure}

Figure~\ref{fig:training} shows that the training and test losses decrease
similarly up to about 15 epochs, after which a gap between them gradually
opens. Both losses nevertheless continue to decrease up to 50 epochs.
For training efficiency, we use the checkpoint at 4000 optimization steps
(\(\sim 12.5\) epochs), corresponding to about one quarter of the full
training run, for all evaluations in this work. All multigrid results below
are thus obtained with a relatively short training, which also reduces the
one-time training cost shared by all configurations of the ensemble.

\section{Multigrid case study}
\label{sec:mg}

DD-$\alpha$AMG~\cite{Frommer:2013} consists of a setup phase and a solve phase. The setup phase constructs the prolongator \(P\) from \(K\) test vectors approximating the near-null space of \(D[U]\) and forms the coarse operator \(D_c=P^\dagger D P\); the solve phase uses a \(V\)-cycle as a preconditioner for FGMRES. For each gauge configuration, the adaptive setup starts from \(K\) random vectors, applies SAP smoothing, constructs \(P\) and \(D_c\) through aggregation, and iteratively refines the test vectors using \(V\)-cycles.

In our test we replace the random vectors by the network outputs
$\{v_k\}_{k=1}^{K}$, smooth them with five SAP iterations, keep
the aggregation, and omit the iterative refinement of the test vectors using $V$-cycles. After setup, we solve the system using FGMRES with the same solver settings for both methods. Thus, only the test-vector setup differs. We
benchmark four models trained for 4000 steps at $\kappa=0.158$ with
$K\in\{12,24,48,64\}$ on two held-out configurations (A and B). The same
models are applied without retraining at $\kappa=0.157$ and near $\kappa_c$.
Since $\kappa_c$ differs from configuration to configuration, we determine
it separately for each one. We compute the low-lying eigenvalues of $D[U]$
and tune $\kappa$ such that the eigenvalue of smallest magnitude is close to zero, which
gives $\kappa=0.15775$ for A and $\kappa=0.1586$ for B. Each configuration
is thus tested close to the point where $D[U]$ becomes singular.
We solve to residual $10^{-10}$, count fine ($D$) and coarse ($D_c$)
operator applications, and record the DD-$\alpha$AMG setup wall time per
configuration. All computations in this work, including the training, are
performed on a single NVIDIA A100 80GB PCIe GPU.

\section{Results}

\begin{figure}[t]
\centering
\includegraphics[width=\textwidth]{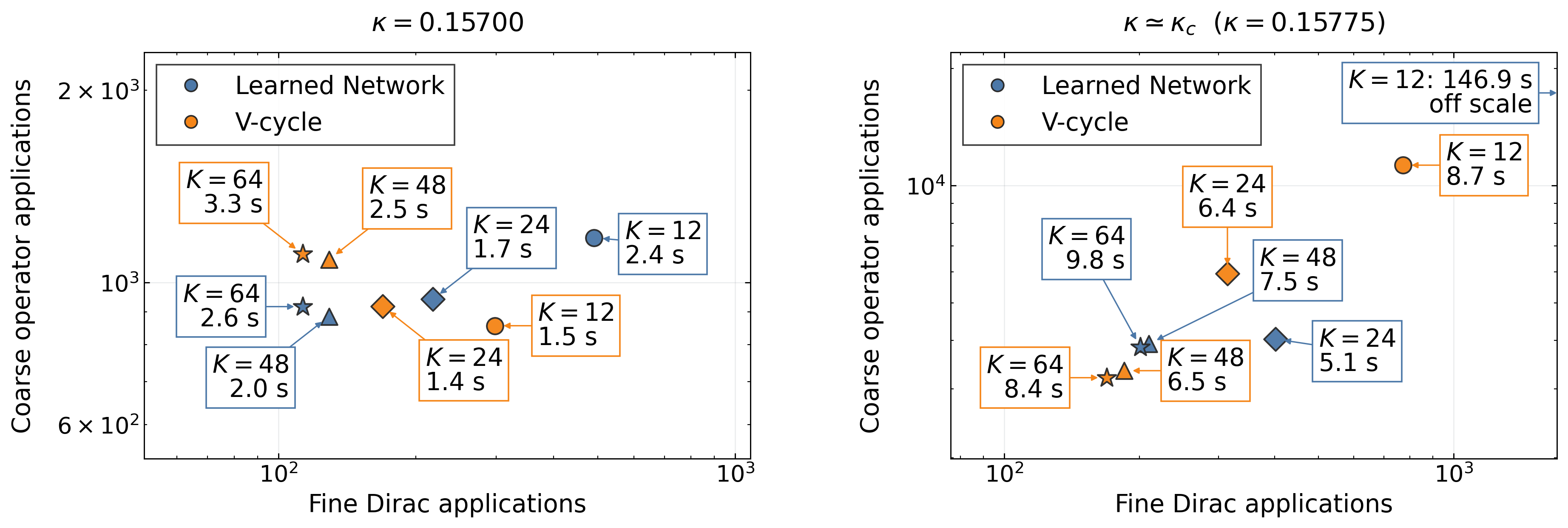}
\caption{Learned vs.\ conventional MG test-vector setup on configuration A
($16^3{\times}32$). Left: $\kappa{=}0.157$; right: near $\kappa_c$.
Axes: fine and coarse operator applications during FGMRES to residual
$10^{-10}$; marker shape indicates $K$ and labels give solve times.}
\label{fig:cfgA}
\end{figure}

\begin{figure}[t]
\centering
\includegraphics[width=\textwidth]{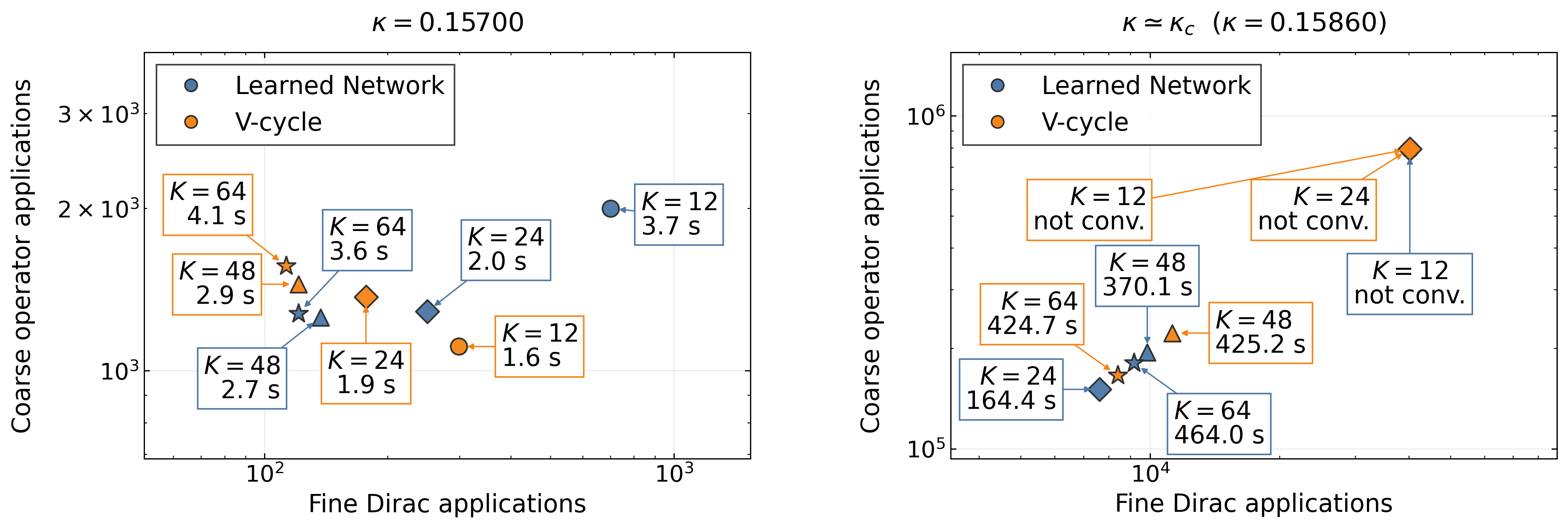}
\caption{Same comparison as Fig.~\ref{fig:cfgA} on configuration B. ``not conv.'' marks runs that did not reach the tolerance within 5000 outer FGMRES iterations.}
\label{fig:cfgB}
\end{figure}

Figures~\ref{fig:cfgA} and~\ref{fig:cfgB} compare the learned setup with the
conventional DD-$\alpha$AMG setup~\cite{Frommer:2013} on configurations A and B
at $\kappa=0.157$ and near $\kappa_c$.
At $\kappa=0.157$, the learned vectors give shorter solve times than the
conventional setup for $K=48$ and $64$ on both configurations, similar times
for $K=24$, and longer times for $K=12$.
Near $\kappa_c$, the learned vectors shorten the solve for $K=24$ on both
configurations. On configuration B, the conventional setup with $K=24$ does
not reach the target residual within 5000 FGMRES iterations (200 restart cycles of length 25), whereas the learned
vectors do. For $K=48$ and $64$ near $\kappa_c$, the solve times of the two
setups differ by less than $20\%$.
For $K=12$, the learned vectors need far more operator applications than the
conventional setup on configuration A, and neither method converges on
configuration B.

\begin{table}[!htb]
\centering
\caption{DD-$\alpha$AMG setup wall time in seconds on configurations~A and B
($16^3{\times}32$), measured on a single NVIDIA A100 80GB PCIe GPU. Near $\kappa_c$ corresponds to $\kappa=0.15775$ (A) and
$\kappa=0.1586$ (B). The one-time network training is excluded.\newline
(Net.): network evaluation and five SAP iterations.\newline
($V$-cyc.): random vectors, three SAP iterations, and six adaptive $V$-cycle iterations.\newline
Both setups end with chiral coarsening and coarse-operator construction.}
\label{tab:setup}
\begin{tabular}{@{}rrrrrrrrr@{}}
\toprule
& \multicolumn{4}{c}{$\kappa=0.157$} & \multicolumn{4}{c}{near $\kappa_c$} \\
\cmidrule(lr){2-5}\cmidrule(lr){6-9}
& \multicolumn{2}{c}{A} & \multicolumn{2}{c}{B}
& \multicolumn{2}{c}{A} & \multicolumn{2}{c}{B} \\
\cmidrule(lr){2-3}\cmidrule(lr){4-5}\cmidrule(lr){6-7}\cmidrule(lr){8-9}
$K$ & Net. & $V$-cyc. & Net. & $V$-cyc. & Net. & $V$-cyc. & Net. & $V$-cyc. \\
\midrule
12 &  5.6 &   8.2 &  5.6 &   8.5 &  5.5 &  10.0 &  5.5 &  11.3 \\
24 &  7.9 &  21.6 &  7.9 &  24.3 &  7.9 &  29.1 &  7.8 &  31.9 \\
48 & 13.7 &  83.3 & 13.7 &  98.8 & 13.7 & 106.0 & 13.6 & 109.1 \\
64 & 18.6 & 156.6 & 18.5 & 183.2 & 18.4 & 188.0 & 18.5 & 204.3 \\
\bottomrule
\end{tabular}
\end{table}

Table~\ref{tab:setup} reports the wall time of the DD-$\alpha$AMG setup on configurations~A and B.
Figures~\ref{fig:cfgA} and~\ref{fig:cfgB} count operator applications in the FGMRES solve phase; the table measures only the setup phase that builds the prolongator and coarse operator.
The network setup, including the network evaluation, takes $5$--$19$\,s and grows slowly with $K$, whereas the conventional $V$-cycle setup grows from about $8$\,s to about $200$\,s.
The setup time is thus reduced by a factor that increases with $K$, from about $1.5$--$2$ at $K=12$ to $8$--$11$ at $K=64$.
When the time of one solve is added to the setup time, the learned approach is faster in every case in which both methods converge, except for $K=12$ near $\kappa_c$ on configuration A.
Training the network for 4000 steps takes about 19--25 h per model.
Although this training cost is substantial, it is incurred only once
for each value of $K$, after which the trained model can be reused
across gauge configurations without retraining.

\section{Summary}

We trained gauge-equivariant networks with a Rayleigh-Ritz loss to produce
approximate low modes of $D^\dagger D$, where $D$ is the Wilson Dirac operator. Within a quenched
ensemble, the Rayleigh-Ritz loss generalizes to held-out configurations. As a case
study, the network outputs can replace the adaptive $V$-cycle test-vector
loop in two-level multigrid. This lowers the setup time per configuration by
a factor that grows with $K$. For $K\geq24$ the solve time stays comparable to
or shorter than with the conventional setup, whereas for $K=12$ the learned
vectors are less effective, in particular near $\kappa_c$. The same framework can be trained with other
lattice Dirac operators. Future work includes extending the study to other ensembles and
larger volumes, as well as further lattice QCD applications beyond
the MG benchmark presented here. We also plan to investigate how
 power series in hopping operators act on the eigenvectors of the Dirac operator
and how the choice of parallel-transport paths affects the learned
low-mode subspace.

\acknowledgments
We thank Akio Tomiya for fruitful discussions. This work is supported by JSPS KAKENHI Grant No. 22H05112. Numerical work was
performed on GPU resources at the Center for Computational Sciences,
University of Tsukuba.

\end{document}